\documentclass[apjl]{emulateapj}
\bibpunct{(}{)}{;}{a}{}{,} 
\usepackage{graphicx}
\usepackage{threeparttable}
\usepackage{natbib,twoopt}

\shorttitle{HR8799 characterization with HST}
\shortauthors{Rajan et al.}

\begin{document}

\title{Characterizing the Atmospheres of the HR8799 Planets with HST/WFC3 }

\author{Abhijith Rajan\altaffilmark{1}, Travis Barman\altaffilmark{2}, R\'emi Soummer\altaffilmark{3}, J. Brendan Hagan\altaffilmark{3,4}, Jennifer Patience\altaffilmark{1}, Laurent Pueyo\altaffilmark{3}, \'Elodie Choquet\altaffilmark{3}, Quinn Konopacky\altaffilmark{5}, Bruce Macintosh\altaffilmark{6}, Christian Marois\altaffilmark{7}}
\altaffiltext{1}{School of Earth and Space Exploration, Arizona State University, AZ 85282}
\altaffiltext{2}{Department of Planetary Sciences and Lunar and Planetary Laboratory, University of Arizona, AZ 85721}
\altaffiltext{3}{Space Telescope Science Institute, Baltimore, MD 21218}
\altaffiltext{4}{Purdue University, West Lafayette IN, 47907}
\altaffiltext{5}{Dunlap Institute for Astronomy \& Astrophysics, University of Toronto, 50 St. George Str. Toronto, ON M5S 3H4, Canada}
\altaffiltext{6}{Kavli Institute for Particle Astrophysics and Cosmology, Stanford University, Stanford, CA 94305, USA}
\altaffiltext{7}{National Research Council of Canada Herzberg, Victoria, BC V9E 2E7, Canada}

\begin{abstract}
We present results from a \emph{Hubble Space Telescope} (\emph{HST}) program characterizing the atmospheres of the outer two planets, in the HR8799 system. The images were taken over 15 orbits in three near-infrared medium-band filters -- F098M, F127M and F139M -- using the Wide Field Camera 3. One of the three filters is sensitive to water absorption band inaccessible from ground-based observations, providing a unique probe of the thermal emission from the atmospheres of these young giant planets. The observations were taken at 30 different spacecraft rolls to enable angular differential imaging, and the full data set was analyzed with the Karhunen-Lo{\'e}ve Image Projection (KLIP) routine, an advanced image processing algorithm adapted to work with \emph{HST} data. To achieve the required high contrast at sub arcsecond resolution, we utilized the pointing accuracy of \emph{HST} in combination with an improved pipeline designed to combine the dithered, angular differential imaging data with an algorithm designed to both improve the image resolution and accurately measure the photometry. The results include F127M ($J$) detections of the outer planets, HR8799~\textit{b} and \textit{c} and the first detection of HR8799b in the water-band (F139M) filter. The F127M photometry for HR8799c agrees well with fitted atmospheric models resolving a long standing difficulty to model the near-IR flux for the planet consistently.
\end{abstract}


\section{Introduction}

Within the past decade, direct imaging of exoplanets has advanced from initial discoveries of favorable low contrast, planetary mass companions like 2M1207b \citep{Chauvin:2005a} and AB~Pic~b \citep{Chauvin:2005b} to high-contrast exoplanets around bright young stars like the HR8799 planetary system \citep{Marois:2008, Marois:2010b}, $\beta$~Pic~b \citep{Lagrange:2010}, and HD~95086b \citep{Rameau:2013}, and GJ~504b \citep{Kuzuhara:2013}. Directly imaged exoplanets form a critical subset of the exoplanet population for which it is possible to characterize atmospheric properties from thermal emission. Given the steep, monotonic decline in planet brightness with time \citep[e.g.][]{Burrows:2001}, the currently known directly imaged exoplanets (companions to young stars), enable investigations into the early evolution of exoplanets. Imaged young atmospheres also present a valuable comparison to the transmission\citep[e.g.][]{Deming:2005, Sing:2008} and emission \citep[e.g.][]{Charbonneau:2005, Knutson:2007} spectra of intensely irradiated planets orbiting older stars.

\emph{HST} has been employed successfully in characterizing exoplanet systems for over a decade \citep{Charbonneau:2002, Soummer:2012}. We present Wide Field Camera 3 imaging of the multiple planet system, HR8799. The system has four young, massive planets ($\sim$4--7 M$_{\rm Jup}$) orbiting a $\sim$30~Myr \citep{Zuckerman:2011} A5 star, amenable to both photometric and spectroscopic follow-up. This is the first time that planets with high contrasts have been successfully imaged with WFC3/IR, a camera without a coronagraph and undersampled point spread function (PSF). The techniques described in the paper should enable future studies of planetary systems with \emph{HST}. In this paper, we characterize the outer two planets, HR8799~\textit{b} and \textit{c}, by combining the new space-based near-infrared photometric data to the existing suite of ground-based data available for the two planets and discuss the implications of the new photometry. 

\begin{figure*}
\centering
\includegraphics[trim=4cm 8.5cm 15cm 4cm, width=14 cm]{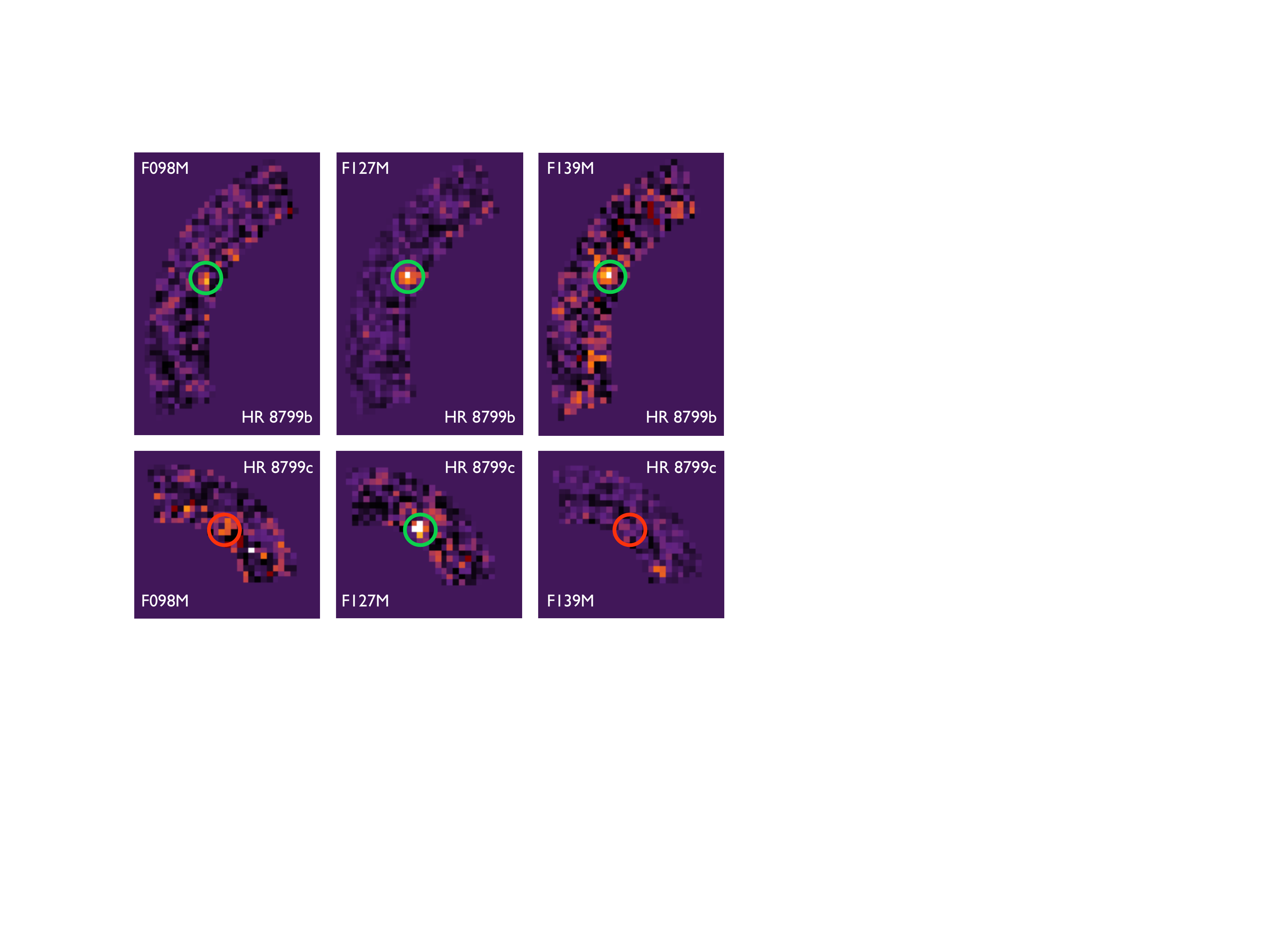}
\vspace{-0.2cm}
\caption{ \label{fig:Detection} Multi-wavelength \emph{HST} HR8799 data. Top: Image zones with locally optimized KLIP detections of HR8799b in all three filters - F098M, F127M and F139M, respectively. Bottom: Same as above, for HR8799c. HR8799c is only detected in the F127M filter, with upper limits in F098M and F139M.}
\end{figure*}

\section{Observations}

Observations for this study\footnote{GO Program \#12511} were taken from Nov 2011 to Dec 2012 with the near-IR channel of WFC3 on the \emph{HST}. Due to the brightness of the primary star (J=5.3~mag), all combinations of sub-array and detector readout time results in saturated zeroth read. In the absence of a coronagraph we designed the observations to saturate the central $\sim$0.$\arcsec5$ of the stellar PSF using 2.2s long observations, focusing the program on the \textit{b} and \textit{c} planets.

The HR8799 system was observed in three medium-band filters, F098M, F127M, and F139M over 15 orbits. The orbits were grouped contiguously in sets of three to increase PSF correlation. The observing sequence was designed to maximize the total roll angle and minimize variations in the PSF of the observations over the duration of the program. To maximize the rotation, the star was observed over two different roll angles in each orbit, resulting in observations taken over 30 separate roll angles covering $\sim$270 degrees (not uniformly sampled) to enable Angular Differential Imaging (ADI) reduction \citep{Marois:2006}. Within each roll, the telescope was dithered using a customized nine-point spiral dither pattern covering a 0.$\arcsec$13$\times$0.$\arcsec$13 region, with half-pixel dithers ($\sim$64~mas), ensuring a well sampled PSF. This observing procedure enhanced stability, but resulted in reduced total exposure time on the object and a limited range of roll angles within each block of three orbits. At the conclusion of the observing program, the full dataset comprised 270 images in each of the three filters.

\section{Data Calibration}
The WFC3/IR images are undersampled data at our wavelengths of interest. The \texttt{DRIZZLEPAC} software is the standard tool to improve the resolution of \emph{HST} data, but it not optimal for high contrast imaging since it corrects the geometric distortion prior to improving the resolution of the images, the opposite being more appropriate here. We therefore modified the \textit{ALICE} pipeline \citep{Soummer:2014,Choquet:2014} to include the calibration of dithered WFC3 images. 

\begin{figure}[!hb]
\centering
\includegraphics[width=\linewidth]{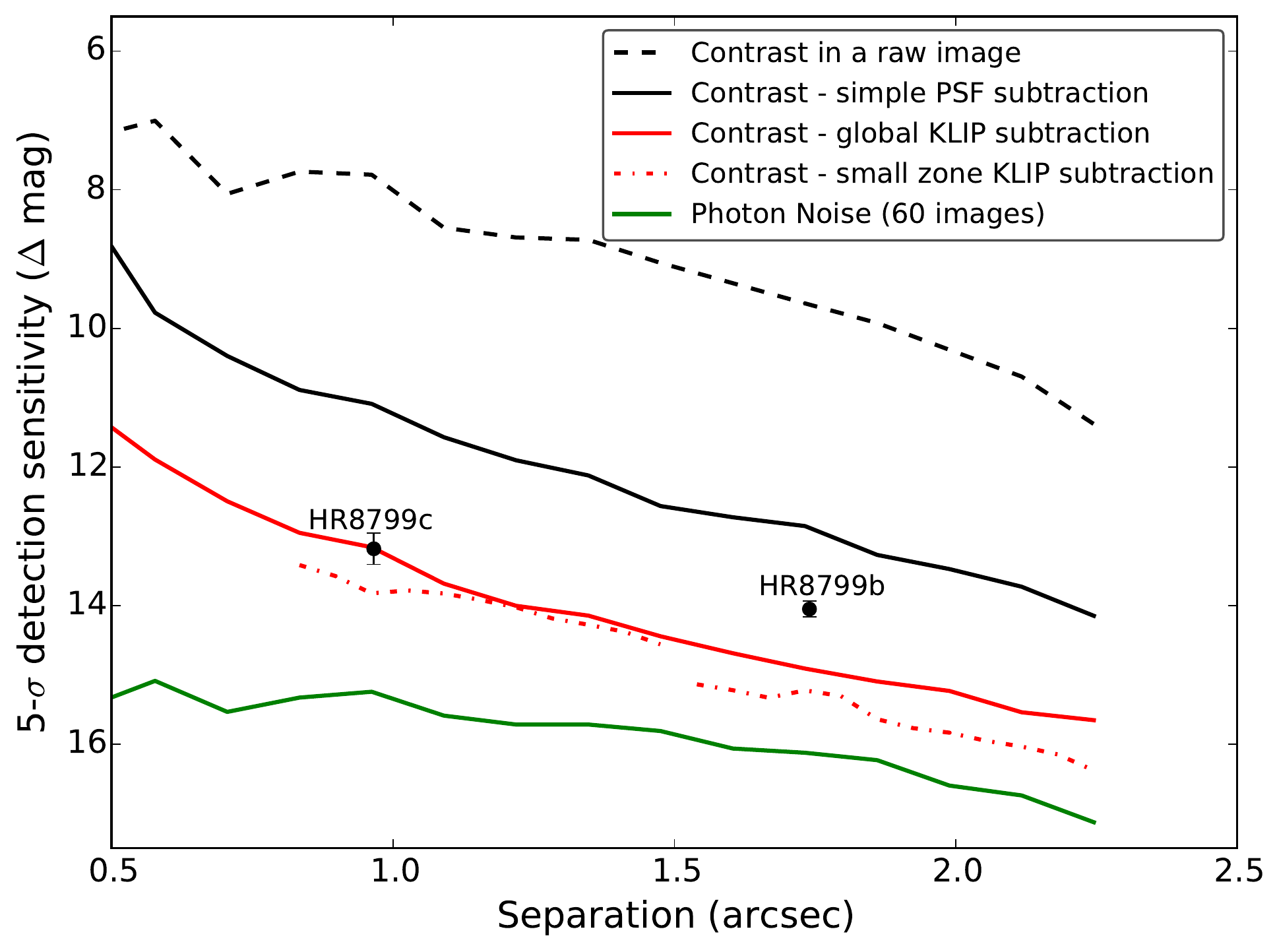}
\vspace{-0.2cm}
\caption{ \label{fig:Contrast} 5-$\sigma$ HR8799 detection sensitivity with $\Delta$F127M magnitudes as a function of angular separation from the star. The black circles are our photometry values for HR8799bc. The solid red line shows the contrast estimated from a global KLIP reduction and the red dashed segments are contrast levels obtained in the optimized zones. This figure is intended to help preparation of future observations at high-contrast with \emph{HST/WFC3} with the green line indicating the photon noise floor for the full dataset.}
\end{figure}

The program used a 9-pt spiral dither pattern with 0.5 pixel dither steps. \emph{HST} pointing stability within a single orbit for small dithers is $\sim$2--5~mas. Using the WCS information and the ``interleave method'' \citep{Lauer:1999}, the dithered frames were combined to improve the resolution by a factor of 2 ($\sim$64~mas/pixel). The selected dither pattern does not permit reconstruction of two fully independent images using interleaving and one dithered frame has to be shared between them. The additional time-sampling provided by generating two images per 9-pt dither meant a deeper contrast and was thus preferred over the one image per 9-pt dither case. The upsampled images are then corrected from the detector distorted frame to an undistorted frame using correction maps (J. Anderson, private comm). Finally, the images were aligned by cross-correlating on the diffraction spikes in the data, resulting in 60 aligned and upsampled images.

The data were reduced by the Karhunen-Lo{\'e}ve Image Projection (KLIP) algorithm based on principal component analysis \citep[PCA,][]{Soummer:2012} used in the ALICE pipeline. For each of the 60 upsampled images, the reference PSF library is assembled imposing a minimum rotation of 2$\times$FWHM at planet location to reduce self-subtraction.

We assumed knowledge of the location of the planets and reduced data in local zones around the expected positions for planets~\textit{b} and \textit{c}, similar to what was done in \citet{Soummer:2011}. The reduction was performed in annular sections, over a parameter space exploring two 
radial sizes and three 
azimuthal widths. With $\sim$25 
KLIP reductions per image our parameter space includes $\sim$150 
images, giving $\sim$9000 images for each reduction of a given planet/filter combination. 
We explored a number of geometries and in particular the location of the planet with respect to the zone with best results when the planet is close to the inner edge of the zone.

Each reduced image is corrected for the KLIP throughput loss using forward modeling, estimated by projecting a \emph{TinyTim} model PSF \citep{Krist:2011} onto the KL modes, as described in \citet{Soummer:2012,Pueyo:2015}. We then compute a data quality criterion as the ratio of the algorithm throughput to a noise estimate with matched-filtering in a local region around the planet. This criterion is therefore proportional to the true SNR, but without introducing any of the planet signal in the calculation so that it does not bias the results by amplifying speckles. The reduced data are ordered according to this criterion and cumulatively-combined using a median. The data quality criterion is then recalculated to determine the optimal number of images in the final image, shown in Fig.~\ref{fig:Detection}. This approach permits exploration of any algorithm parameter space to produce a single final image automatically by identifying the best combination of algorithm throughput and speckle suppression. 

Photometry was obtained using matched-filtering with a truncated model PSF (5 pix) to reduce potential contamination from local speckles. This matched filter combined with partial truncation of the planet PSF by the reduction zone leads to an incomplete fraction  of the planet flux $(\sim 80\%)$. All these effects were carefully calibrated using aperture photometry on archival WFC3 data of white dwarfs (GD153, G191B2B) from \emph{HST} calibration programs, and of similar spectral type isolated brown dwarfs. The overall photometric correction precision is of the order of 3-4\%, significantly smaller than the final photometric error bars on the planets. 

The PSF library is not rigorously free of companion contribution, however the contribution of the companion at other roll angles is limited, since the data is not a true ADI sequence (the orients span the 270 degrees over multiple epochs) and the planets are very faint compared to the PSF wings. In addition, the noise is zero-mean within the reduction zone from KLIP but is not necessarily zero-mean within the matched-filter equivalent aperture, and small biases may remain. The validity of the forward modeling approach for this particular dataset was investigated by injecting \emph{TinyTim} PSFs in the data at the planets' radial separation over a range of azimuthal angles (25 PSFs for \textit{b} and 20 PSFs for \textit{c}). The injected PSF flux was iteratively adjusted to result in the same SNR as each true planet after reduction. The mean photometric error on the synthetic planets was used as a bias correction to account for the two effects discussed above and the error on the photometry was estimated from the standard deviation in the measured signal for each of the fakes. 
Upper limit detections for HR 8799 \textit{c} in F098M and F139M were obtained by adjusting the injected PSFs to detect almost all of them. 
To ensure the absence of speckle amplification we injected ``zero-flux synthetic planets'' using the exact same pipeline. No significant detection could be noticed at the location of these injected zero-flux PSFs. Various combinations of reduced images were tested for the presence of flux at given positions as a function of telescope roll angle and reduction parameters, see \citet{Soummer:2011} for a detailed description. We estimated the False Positive Fraction (FPF) of 7.6$\times10^{-22}$ (SNR $\sim$10) and 3.8$\times10^{-7}$ (SNR$\sim$5.3) for b and c in the F127M filter assuming Gaussian statistics. We also calculated an FPF of 2.7$\times10^{-11}$ (SNR$\sim$7) and 1.3$\times10^{-4}$ (SNR$\sim$4) for b and c in F127M filter assuming a small number of resolution elements using the Student's t-test relation presented in \citet{Mawet:2014}.

\section{Results}

\begin{figure}[b]
\centering
\includegraphics[width=\linewidth]{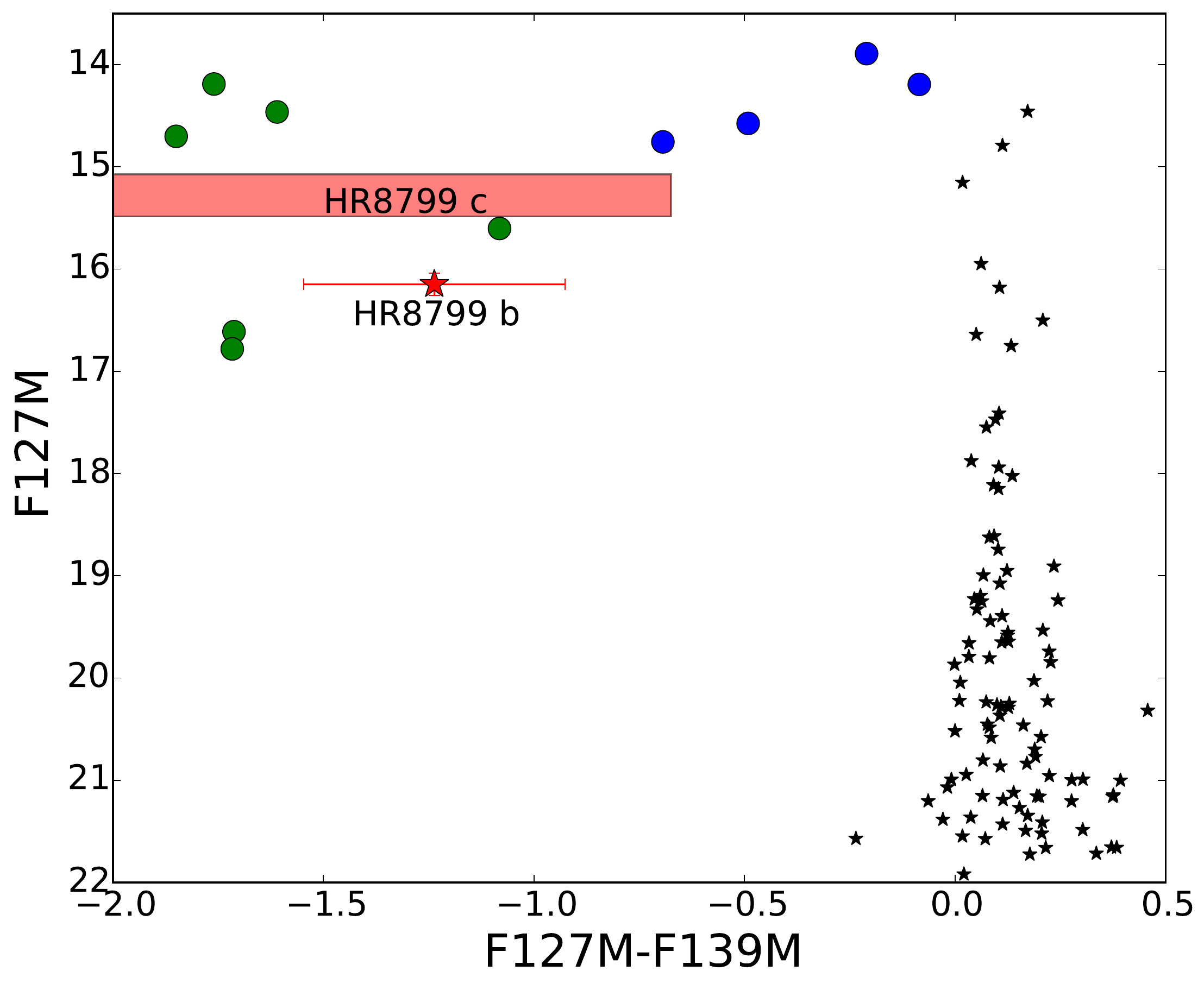}
\vspace{-0.2cm}
\caption{ \label{fig:CMD} \emph{HST} near-infrared color-magnitude diagram. The HR8799bc planets (red star, shaded region) are plotted with field L (blue circles) and T (green circles) brown dwarfs from recent \emph{HST} studies \citep{Aberasturi:2014, Apai:2013,Yang:2014}. The black points are background field targets. The red rectangle for \textit{c} represents our upper-limit estimate.}
\end{figure}

Contrast curves for the data in the F127M filter are shown in Fig.~\ref{fig:Contrast}, the figure indicates the contrast achieved using different analysis techniques and the contrast in the global-KLIP and small-zone KLIP reductions. The global-KLIP reduction is not used in our analysis, we present this contrast curve estimate to guide future possible studies at high-contrast with \emph{HST/WFC3-IR}. The measured \texttt{VEGAMAG} photometry for the detections and 1$\sigma$ limits is presented in Table~\ref{Tab:Phot} using revised zero-points\footnote{Revised ZP: F098M = 24.2209 mag, F127M = 23.7503 mag, F139M = 23.7679 mag}. The photometry of the planets is compared with brown dwarfs measured in the same \emph{HST} filters in the color-magnitude diagram in Fig.~\ref{fig:CMD}. The positions of the planets in Fig.~\ref{fig:CMD}, verifies that the flux measurement for the planets do not deviate unrealistically from the brown dwarfs. The spectral type derived for each planet is dependent on the contrast between $J$ and water band fluxes being similar to that measured for brown dwarfs. Using a relation converting \emph{HST} photometric colors to a spectral type from \citet{Aberasturi:2014}\footnote{SpT=1.56-6.25*(F127M-F139M)}, we calculate a spectral type of L9.5 $\pm$ 0.5 for HR8799b and $>$L7 for HR8799c. The errors are the quadrature sum of the intrinsic scatter in the relation and the error in the photometric color. The larger role of clouds in giant planets with temperatures comparable to cloud-free brown dwarfs complicates the conversion of the color-magnitude position into spectral types. The following two subsections model the combined \emph{HST} photometry with measurements from the literature at different wavelengths to determine the atmospheric properties.

\subsection{HR8799b}

The \emph{HST} data in F139M provide the first detection of the planet \textit{b} in the water absorption region at 1.4$\micron$. Additionally, the photometric points in F098M and the peak of the $J$ band (F127M) are consistent with measurements taken across similar wavelengths with ground based instruments \citep{Currie:2011, Marois:2008, Oppenheimer:2013}. We combine the new \emph{HST} data with photometric data available in the literature and fit the fluxes to synthetic model spectra from \citet{Barman:2011, Barman:2015} (cited as B15 in the rest of the paper). The models include clouds located at the intersection of the pressure-temperature profile and the chemical equilibrium condensation curve with a parameterized thickness but a homogeneous distribution in latitude and longitude across the planet. The models also include non-equilibrium chemistry for all important C, N and O bearing molecules, as well as updated line lists for CH4 and NH3 (see B15 for details). The particle size-distribution is centered on 5$\micron$ following a log-normal distribution.

\begin{figure*}
\centering
\includegraphics[width=\textwidth]{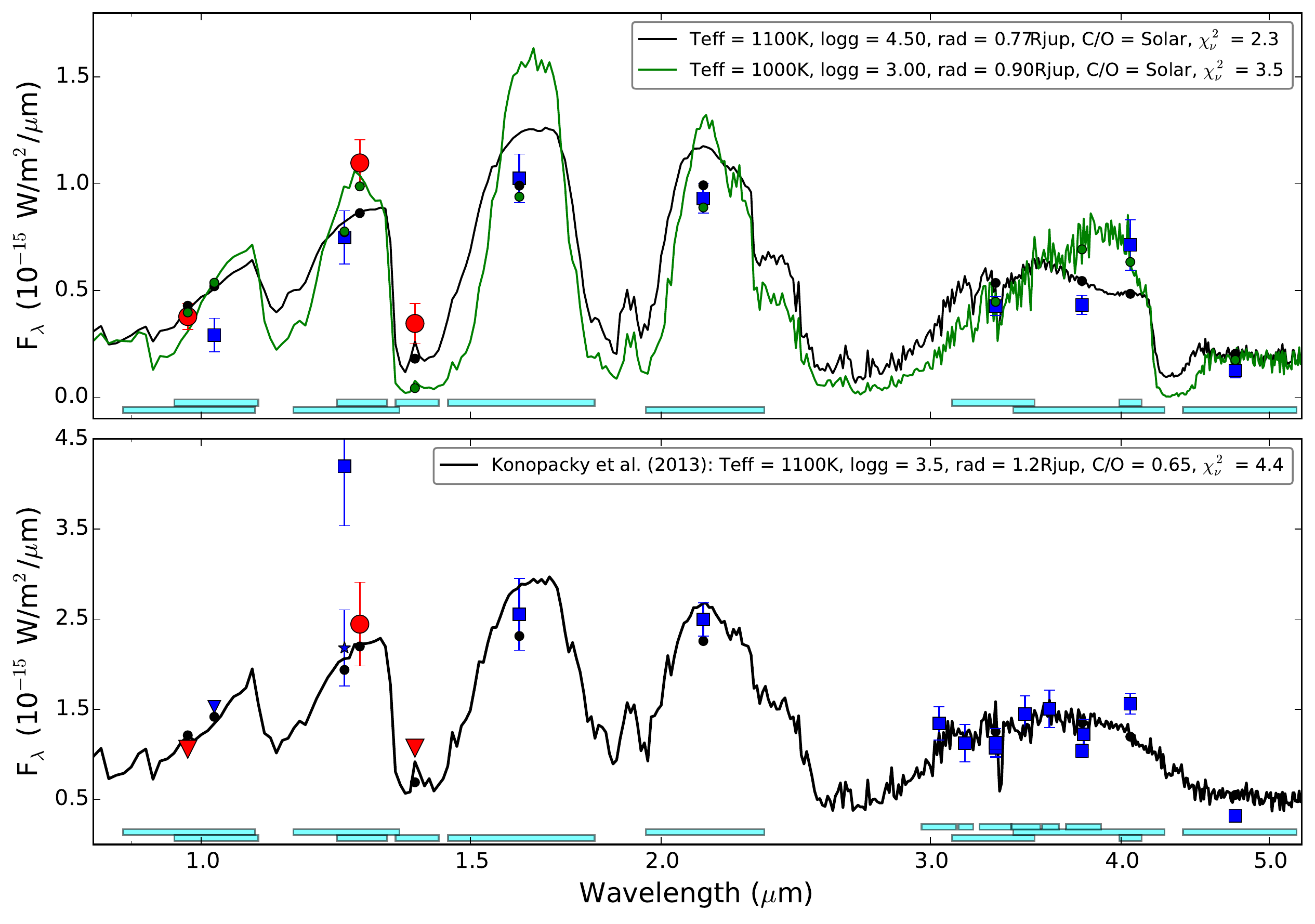}
\vspace{-0.2cm}
\caption{ \label{fig:BestFitModel} Top: Model fits to HR8799b data. In the figure the red circles are \emph{HST} photometry, and the blue squares are ground-based photometry, with the filter width shown as cyan bars at the bottom of the plot. The black and green lines (and corresponding circles) are the model spectra fitted to the full dataset. Bottom: The black line is the \citet{Konopacky:2013} HR8799c model. The ground-based photometry in this paper comes from \citet{Marois:2008,Currie:2011, Oppenheimer:2013, Currie:2014, Skemer:2012, Skemer:2014, Galicher:2011}. The blue star for HR8799c is the $J$-band photometry from \citet{Oppenheimer:2013}. }
\end{figure*}

Figure \ref{fig:BestFitModel} (top) shows two fits for the HR8799b photometry, the best fitting model to the F139M (black line), and the best fit to the [4.05] point \citep[green line]{Currie:2014}. No individual model was able to fit the pair of neighboring fluxes at either F127M and F139M or $L^\prime$ and [4.05]. The discrepancy may indicate missing physical processes in the model and the data is complicated by the non-contemporaneous measurements. Although the [4.05] band covers the Br-$\alpha$ line, accretion is unlikely due to the lack of Br-$\gamma$ emission in the K-band spectrum \citep{Barman:2011}. The ratio of F139M flux to F127M flux may be useful for estimating the combined effects of cloud size distribution, cloud thickness and coverage. However, effective temperature, gravity and composition will need to be better constrained to avoid degeneracies in the model fitting and better agreement with evolutionary models.

Fitting a combination of near-IR spectroscopy and IR photometry, B15 found a best matching model with T$_{\rm eff}$ = 1000K and log(g) = 3.5, with potentially subsolar water abundance and enhanced C/O, however, the uncertain surface gravity results in a wide spread of values including solar metallicity.  In this paper, the cooler best-fit model has solar abundances and we find T$_{\rm eff}$ = 1000--1100~K and log(g) = 3.0--4.5, indicating consistency between photometry-based fits and those including spectroscopy. Our measured water band flux, relative to $J$-band, however, is brighter than predicted by all of the non-solar models explored by B15, suggesting that perhaps clouds play an important role.

\subsection{HR8799c}

For HR8799c we measured the photometry in the F127M band and upper limits were determined for the F098M and F139M filters. The F127M flux is consistent with values obtained by integrating the spectrum from P1640 over the same wavelength range \citep{Oppenheimer:2013}, but substantially lower than the reported $J$-band photometry \citep{Marois:2008}. The new \emph{HST} data were combined with similar and longer wavelength fluxes from the literature to construct the spectral energy distribution given in Fig.~\ref{fig:BestFitModel} (bottom). Plotted with the data is a model with T$_{\rm eff}$ = 1100K, log(g) = 3.5 and elevated C/O that fit the K-band spectrum from \citet{Konopacky:2013}. The fits agree well with the \emph{HST} photometry and the longer wavelength data with the exception of a significant difference with the earliest measurement of the $J$-band photometry and an $\sim$2.5$\sigma$ difference with the [4.05] data point. Our models for HR8799c with homogeneous cloud model agree, within the uncertainties, with the \emph{HST} photometry and longer wavelength data out to $\sim$3.5$\micron$ with the exception of the first epoch of $J$-band photometry\footnote{Improved analysis techniques and stellar variability cannot account for the factor of two difference in flux}. The difficulty in fitting all available photometry with reported uncertainties persists even when using patchy cloud models formed by combining clouds at different temperatures \citep{Skemer:2014} or opacity \citep{Currie:2014} and the \emph{HST} F127M measurement is more consistent with the patchy cloud models. Our attempts with a linear combination of models did not provide a better fit than the homogeneous cloud model with higher C/O ratio presented in Fig.~\ref{fig:BestFitModel} (bottom).

\begin{table}
    \small
    \centering
    \caption{HR8799bc absolute photometry} \label{Tab:Phot}
    \begin{threeparttable}
        \begin{tabular}{|c|ccc|}
            \hline
            Planet & \multicolumn{3}{|c|}{HST Photometry} \\
                   & F098M &  F127M\tnote{a} & F139M \\
            \hline \hline
            HR8799b & 16.90 $\pm$ 0.18 & 16.20 $\pm$ 0.12 & 17.36 $\pm$ 0.26 \\
            HR8799c & $>$15.72     & 15.38 $\pm$ 0.17 & $>$16.16     \\
            \hline
        \end{tabular}
        \begin{tablenotes}
            \item[a]{P1640 spectrum through the F127M filter gives, HR8799b = $16.16\pm0.27$ and HR8799c = $15.17\pm0.08$.}
        \end{tablenotes}
    \end{threeparttable}
\end{table}

\section{Discussion}

Among the \emph{HST} filters, the F127M is the most analogous to a ground-based filter and can be directly compared to previous results reported in the $J$-band. The HR8799b F127M photometry is consistent with previous results \citep{Marois:2008}. There are two prior reported values for HR8799c - a P1640 spectrum \citep{Oppenheimer:2013} and a Keck $J$-band flux \citep{Marois:2008}. The ground-based $J$-band photometry for HR8799c is approximately twice as bright as the measured \emph{HST} flux. Integrating the HR8799c flux calibrated P1640 spectrum through the F127M filter matches the \emph{HST} photometry to within 2-$\sigma$. A potential solution for this discrepancy might be intrinsic photometric variability caused by heterogeneous cloud layers, which we find is not required in our model fits. The early groundbased photometry might also have suffered from calibration issues which combined by the intrinsic variability of the star might explain some of the difference in flux measurement.

Efforts to match synthetic spectra to the ensemble of photometric data for HR8799b result in high effective temperatures and corresponding radii that are smaller by $\sim$50\% (0.69 - 0.92 R$_{\rm Jup}$), than predicted by theoretical brown dwarf and giant planet cooling tracks \citep[see][for a summary]{Marley:2012}. Difficulty finding a model spectrum that simultaneously matches the near and mid-IR photometry could be due to the non-contemporaneous nature of the observations or perhaps the model-observation inconsistencies at multiple bands are an indication of large-scale flux variations. Large variability would bias such model comparisons. For example, the bright $J$-band flux from \citet{Marois:2008} is more consistent with higher effective temperatures than our \emph{HST} F127M flux. The deep water absorption demonstrates that the atmospheres of both \textit{b} and \textit{c} are not enshrouded in high altitude hazes or clouds many pressure scale heights thick, important for many transiting exoplanets \citep{Kreidberg:2014}. Nonetheless, clouds are important in shaping the overall SED of the planets. Both best-matching models plotted in Fig.~\ref{fig:BestFitModel} (top panel) have clouds composed primarily of Iron and Magnesium-Silcate grains, located in the near-infrared photosphere. The cooler model has a cloud located at Pgas $\sim$ 1~bar and extending upward 1 pressure scale-height. The warmer (and higher gravity) model has a cloud base near 10~bar, extending upward 2 pressure scale heights. Using single models to reproduce the observations assumes global cloud coverage which is probably an overestimation. 

With spectral energy distributions well sampled observationally and with model spectra that match reasonably well, the bolometric luminosities of both planets can be estimated, and we determine L$_{\rm bol}$ values of -5.1$\pm$0.1 for \textit{b} and -4.7$\pm$0.1 for \textit{c} consistent with \citet{Marois:2008}.

\section{Conclusions}
In this study, we have demonstrated that WFC3 is capable of investigating the atmospheres of planets requiring high contrast at wavelengths either inaccessible from the ground or at which the extreme adaptive optics systems perform poorly. With its photometric stability, \emph{HST} provides a valuable resource to explore exoplanet atmospheres and will enable measurements such as the search for atmospheric variability. The data from the current program can be used to build a PSF library for future high contrast imaging programs. Based on experience from the current program future studies would benefit from obtaining more exposures rather than performing the time-consuming in-orbit roll.

In this study, we were able to detect HR8799b in three WFC3 filters. In this work and in other studies, matching all of the data simultaneously has not been possible. In particular, some of the shortest wavelength data favor warmer conditions while some of the longer wavelength IR data favor cooler conditions, possibly caused by clouds. For HR8799c, we find very good agreement between the model fit with the F127M, resolving the long standing difficulty of fitting the previously reported $J$-band photometry. A possible explanation for the F127M and $J$-band flux difference is variability. Future spectrophotometric monitoring of HR8799b and HR8799c in the near and mid-IR might resolve whether the planets are variable over the duration of a single rotation period. 

\acknowledgments
Based on observations made with the Hubble Space Telescope, associated with program \#12511. This research was supported in part by NASA cooperative agreements HST-GO-12511.04, HST-AR-12652.01, NSF AST-1411868.


\begin{thebibliography}{37}
\expandafter\ifx\csname natexlab\endcsname\relax\def\natexlab#1{#1}\fi

\bibitem[{{Aberasturi} {et~al.}(2014){Aberasturi}, {Burgasser}, {Mora},
  {Solano}, {Mart{\'i}n}, {Reid}, \& {Looper}}]{Aberasturi:2014}
{Aberasturi}, M., {Burgasser}, A.~J., {Mora}, A., {Solano}, E., {Mart{\'i}n},
  E.~L., {Reid}, I.~N., \& {Looper}, D. 2014, ArXiv e-prints

\bibitem[{{Apai} {et~al.}(2013){Apai}, {Radigan}, {Buenzli}, {Burrows}, {Reid},
  \& {Jayawardhana}}]{Apai:2013}
{Apai}, D., {Radigan}, J., {Buenzli}, E., {Burrows}, A., {Reid}, I.~N., \&
  {Jayawardhana}, R. 2013, \apj, 768, 121

\bibitem[{{Barman} {et~al.}(2015){Barman}, {Konopacky}, {Macintosh}, \&
  {Marois}}]{Barman:2015}
{Barman}, T.~S., {Konopacky}, Q.~M., {Macintosh}, B., \& {Marois}, C. 2015,
  ArXiv e-prints

\bibitem[{{Barman} {et~al.}(2011){Barman}, {Macintosh}, {Konopacky}, \&
  {Marois}}]{Barman:2011}
{Barman}, T.~S., {Macintosh}, B., {Konopacky}, Q.~M., \& {Marois}, C. 2011,
  \apj, 733, 65

\bibitem[{{Burrows} {et~al.}(2001){Burrows}, {Hubbard}, {Lunine}, \&
  {Liebert}}]{Burrows:2001}
{Burrows}, A., {Hubbard}, W.~B., {Lunine}, J.~I., \& {Liebert}, J. 2001,
  Reviews of Modern Physics, 73, 719

\bibitem[{{Charbonneau} {et~al.}(2005){Charbonneau}, {Allen}, {Megeath},
  {Torres}, {Alonso}, {Brown}, {Gilliland}, {Latham}, {Mandushev}, {O'Donovan},
  \& {Sozzetti}}]{Charbonneau:2005}
{Charbonneau}, D., {Allen}, L.~E., {Megeath}, S.~T., {Torres}, G., {Alonso},
  R., {Brown}, T.~M., {Gilliland}, R.~L., {Latham}, D.~W., {Mandushev}, G.,
  {O'Donovan}, F.~T., \& {Sozzetti}, A. 2005, \apj, 626, 523

\bibitem[{{Charbonneau} {et~al.}(2002){Charbonneau}, {Brown}, {Noyes}, \&
  {Gilliland}}]{Charbonneau:2002}
{Charbonneau}, D., {Brown}, T.~M., {Noyes}, R.~W., \& {Gilliland}, R.~L. 2002,
  \apj, 568, 377

\bibitem[{{Chauvin} {et~al.}(2005{\natexlab{a}}){Chauvin}, {Lagrange}, {Dumas},
  {Zuckerman}, {Mouillet}, {Song}, {Beuzit}, \& {Lowrance}}]{Chauvin:2005a}
{Chauvin}, G., {Lagrange}, A.-M., {Dumas}, C., {Zuckerman}, B., {Mouillet}, D.,
  {Song}, I., {Beuzit}, J.-L., \& {Lowrance}, P. 2005{\natexlab{a}}, \aap, 438,
  L25

\bibitem[{{Chauvin} {et~al.}(2005{\natexlab{b}}){Chauvin}, {Lagrange},
  {Zuckerman}, {Dumas}, {Mouillet}, {Song}, {Beuzit}, {Lowrance}, \&
  {Bessell}}]{Chauvin:2005b}
{Chauvin}, G., {Lagrange}, A.-M., {Zuckerman}, B., {Dumas}, C., {Mouillet}, D.,
  {Song}, I., {Beuzit}, J.-L., {Lowrance}, P., \& {Bessell}, M.~S.
  2005{\natexlab{b}}, \aap, 438, L29

\bibitem[{{Choquet} {et~al.}(2014){Choquet}, {Pueyo}, {Hagan}, {Gofas-Salas},
  {Rajan}, {Chen}, {Perrin}, {Debes}, {Golimowski}, {Hines}, {N'Diaye},
  {Schneider}, {Mawet}, {Marois}, \& {Soummer}}]{Choquet:2014}
{Choquet}, {\'E}., {Pueyo}, L., {Hagan}, J.~B., {Gofas-Salas}, E., {Rajan}, A.,
  {Chen}, C., {Perrin}, M.~D., {Debes}, J., {Golimowski}, D., {Hines}, D.~C.,
  {N'Diaye}, M., {Schneider}, G., {Mawet}, D., {Marois}, C., \& {Soummer}, R.
  2014, in Society of Photo-Optical Instrumentation Engineers (SPIE) Conference
  Series, Vol. 9143, Society of Photo-Optical Instrumentation Engineers (SPIE)
  Conference Series, 57

\bibitem[{{Currie} {et~al.}(2014){Currie}, {Burrows}, {Girard}, {Cloutier},
  {Fukagawa}, {Sorahana}, {Kuchner}, {Kenyon}, {Madhusudhan}, {Itoh},
  {Jayawardhana}, {Matsumura}, \& {Pyo}}]{Currie:2014}
{Currie}, T., {Burrows}, A., {Girard}, J.~H., {Cloutier}, R., {Fukagawa}, M.,
  {Sorahana}, S., {Kuchner}, M., {Kenyon}, S.~J., {Madhusudhan}, N., {Itoh},
  Y., {Jayawardhana}, R., {Matsumura}, S., \& {Pyo}, T.-S. 2014, ArXiv e-prints

\bibitem[{{Currie} {et~al.}(2011){Currie}, {Burrows}, {Itoh}, {Matsumura},
  {Fukagawa}, {Apai}, {Madhusudhan}, {Hinz}, {Rodigas}, {Kasper}, {Pyo}, \&
  {Ogino}}]{Currie:2011}
{Currie}, T., {Burrows}, A., {Itoh}, Y., {Matsumura}, S., {Fukagawa}, M.,
  {Apai}, D., {Madhusudhan}, N., {Hinz}, P.~M., {Rodigas}, T.~J., {Kasper}, M.,
  {Pyo}, T.-S., \& {Ogino}, S. 2011, \apj, 729, 128

\bibitem[{{Deming} {et~al.}(2005){Deming}, {Brown}, {Charbonneau},
  {Harrington}, \& {Richardson}}]{Deming:2005}
{Deming}, D., {Brown}, T.~M., {Charbonneau}, D., {Harrington}, J., \&
  {Richardson}, L.~J. 2005, \apj, 622, 1149

\bibitem[{{Galicher} {et~al.}(2011){Galicher}, {Marois}, {Macintosh}, {Barman},
  \& {Konopacky}}]{Galicher:2011}
{Galicher}, R., {Marois}, C., {Macintosh}, B., {Barman}, T., \& {Konopacky}, Q.
  2011, \apjl, 739, L41

\bibitem[{{Knutson} {et~al.}(2007){Knutson}, {Charbonneau}, {Allen}, {Fortney},
  {Agol}, {Cowan}, {Showman}, {Cooper}, \& {Megeath}}]{Knutson:2007}
{Knutson}, H.~A., {Charbonneau}, D., {Allen}, L.~E., {Fortney}, J.~J., {Agol},
  E., {Cowan}, N.~B., {Showman}, A.~P., {Cooper}, C.~S., \& {Megeath}, S.~T.
  2007, \nat, 447, 183

\bibitem[{{Konopacky} {et~al.}(2013){Konopacky}, {Barman}, {Macintosh}, \&
  {Marois}}]{Konopacky:2013}
{Konopacky}, Q.~M., {Barman}, T.~S., {Macintosh}, B.~A., \& {Marois}, C. 2013,
  Science, 339, 1398

\bibitem[{{Kreidberg} {et~al.}(2014){Kreidberg}, {Bean}, {D{\'e}sert},
  {Benneke}, {Deming}, {Stevenson}, {Seager}, {Berta-Thompson}, {Seifahrt}, \&
  {Homeier}}]{Kreidberg:2014}
{Kreidberg}, L., {Bean}, J.~L., {D{\'e}sert}, J.-M., {Benneke}, B., {Deming},
  D., {Stevenson}, K.~B., {Seager}, S., {Berta-Thompson}, Z., {Seifahrt}, A.,
  \& {Homeier}, D. 2014, \nat, 505, 69

\bibitem[{{Krist} {et~al.}(2011){Krist}, {Hook}, \& {Stoehr}}]{Krist:2011}
{Krist}, J.~E., {Hook}, R.~N., \& {Stoehr}, F. 2011, in Society of
  Photo-Optical Instrumentation Engineers (SPIE) Conference Series, Vol. 8127,
  Society of Photo-Optical Instrumentation Engineers (SPIE) Conference Series,
  0

\bibitem[{{Kuzuhara} {et~al.}(2013){Kuzuhara}, {Tamura}, {Kudo}, {Janson},
  {Kandori}, {Brandt}, {Thalmann}, {Spiegel}, {Biller}, {Carson}, {Hori},
  {Suzuki}, {Burrows}, {Henning}, {Turner}, {McElwain}, {Moro-Mart{\'{\i}}n},
  {Suenaga}, {Takahashi}, {Kwon}, {Lucas}, {Abe}, {Brandner}, {Egner}, {Feldt},
  {Fujiwara}, {Goto}, {Grady}, {Guyon}, {Hashimoto}, {Hayano}, {Hayashi},
  {Hayashi}, {Hodapp}, {Ishii}, {Iye}, {Knapp}, {Matsuo}, {Mayama}, {Miyama},
  {Morino}, {Nishikawa}, {Nishimura}, {Kotani}, {Kusakabe}, {Pyo}, {Serabyn},
  {Suto}, {Takami}, {Takato}, {Terada}, {Tomono}, {Watanabe}, {Wisniewski},
  {Yamada}, {Takami}, \& {Usuda}}]{Kuzuhara:2013}
{Kuzuhara}, M., {Tamura}, M., {Kudo}, T., {Janson}, M., {Kandori}, R.,
  {Brandt}, T.~D., {Thalmann}, C., {Spiegel}, D., {Biller}, B., {Carson}, J.,
  {Hori}, Y., {Suzuki}, R., {Burrows}, A., {Henning}, T., {Turner}, E.~L.,
  {McElwain}, M.~W., {Moro-Mart{\'{\i}}n}, A., {Suenaga}, T., {Takahashi},
  Y.~H., {Kwon}, J., {Lucas}, P., {Abe}, L., {Brandner}, W., {Egner}, S.,
  {Feldt}, M., {Fujiwara}, H., {Goto}, M., {Grady}, C.~A., {Guyon}, O.,
  {Hashimoto}, J., {Hayano}, Y., {Hayashi}, M., {Hayashi}, S.~S., {Hodapp},
  K.~W., {Ishii}, M., {Iye}, M., {Knapp}, G.~R., {Matsuo}, T., {Mayama}, S.,
  {Miyama}, S., {Morino}, J.-I., {Nishikawa}, J., {Nishimura}, T., {Kotani},
  T., {Kusakabe}, N., {Pyo}, T.-S., {Serabyn}, E., {Suto}, H., {Takami}, M.,
  {Takato}, N., {Terada}, H., {Tomono}, D., {Watanabe}, M., {Wisniewski},
  J.~P., {Yamada}, T., {Takami}, H., \& {Usuda}, T. 2013, \apj, 774, 11

\bibitem[{{Lagrange} {et~al.}(2010){Lagrange}, {Bonnefoy}, {Chauvin}, {Apai},
  {Ehrenreich}, {Boccaletti}, {Gratadour}, {Rouan}, {Mouillet}, {Lacour}, \&
  {Kasper}}]{Lagrange:2010}
{Lagrange}, A.-M., {Bonnefoy}, M., {Chauvin}, G., {Apai}, D., {Ehrenreich}, D.,
  {Boccaletti}, A., {Gratadour}, D., {Rouan}, D., {Mouillet}, D., {Lacour}, S.,
  \& {Kasper}, M. 2010, Science, 329, 57

\bibitem[{{Lauer}(1999)}]{Lauer:1999}
{Lauer}, T.~R. 1999, \pasp, 111, 1434

\bibitem[{{Marley} {et~al.}(2012){Marley}, {Saumon}, {Cushing}, {Ackerman},
  {Fortney}, \& {Freedman}}]{Marley:2012}
{Marley}, M.~S., {Saumon}, D., {Cushing}, M., {Ackerman}, A.~S., {Fortney},
  J.~J., \& {Freedman}, R. 2012, \apj, 754, 135

\bibitem[{{Marois} {et~al.}(2006){Marois}, {Lafreni{\`e}re}, {Doyon},
  {Macintosh}, \& {Nadeau}}]{Marois:2006}
{Marois}, C., {Lafreni{\`e}re}, D., {Doyon}, R., {Macintosh}, B., \& {Nadeau},
  D. 2006, \apj, 641, 556

\bibitem[{{Marois} {et~al.}(2008){Marois}, {Macintosh}, {Barman}, {Zuckerman},
  {Song}, {Patience}, {Lafreni{\`e}re}, \& {Doyon}}]{Marois:2008}
{Marois}, C., {Macintosh}, B., {Barman}, T., {Zuckerman}, B., {Song}, I.,
  {Patience}, J., {Lafreni{\`e}re}, D., \& {Doyon}, R. 2008, Science, 322, 1348

\bibitem[{{Marois} {et~al.}(2010){Marois}, {Zuckerman}, {Konopacky},
  {Macintosh}, \& {Barman}}]{Marois:2010b}
{Marois}, C., {Zuckerman}, B., {Konopacky}, Q.~M., {Macintosh}, B., \&
  {Barman}, T. 2010, \nat, 468, 1080

\bibitem[{{Mawet} {et~al.}(2014){Mawet}, {Milli}, {Wahhaj}, {Pelat}, {Absil},
  {Delacroix}, {Boccaletti}, {Kasper}, {Kenworthy}, {Marois}, {Mennesson}, \&
  {Pueyo}}]{Mawet:2014}
{Mawet}, D., {Milli}, J., {Wahhaj}, Z., {Pelat}, D., {Absil}, O., {Delacroix},
  C., {Boccaletti}, A., {Kasper}, M., {Kenworthy}, M., {Marois}, C.,
  {Mennesson}, B., \& {Pueyo}, L. 2014, \apj, 792, 97

\bibitem[{{Oppenheimer} {et~al.}(2013){Oppenheimer}, {Baranec}, {Beichman},
  {Brenner}, {Burruss}, {Cady}, {Crepp}, {Dekany}, {Fergus}, {Hale},
  {Hillenbrand}, {Hinkley}, {Hogg}, {King}, {Ligon}, {Lockhart}, {Nilsson},
  {Parry}, {Pueyo}, {Rice}, {Roberts}, {Roberts}, {Shao}, {Sivaramakrishnan},
  {Soummer}, {Truong}, {Vasisht}, {Veicht}, {Vescelus}, {Wallace}, {Zhai}, \&
  {Zimmerman}}]{Oppenheimer:2013}
{Oppenheimer}, B.~R., {Baranec}, C., {Beichman}, C., {Brenner}, D., {Burruss},
  R., {Cady}, E., {Crepp}, J.~R., {Dekany}, R., {Fergus}, R., {Hale}, D.,
  {Hillenbrand}, L., {Hinkley}, S., {Hogg}, D.~W., {King}, D., {Ligon}, E.~R.,
  {Lockhart}, T., {Nilsson}, R., {Parry}, I.~R., {Pueyo}, L., {Rice}, E.,
  {Roberts}, J.~E., {Roberts}, Jr., L.~C., {Shao}, M., {Sivaramakrishnan}, A.,
  {Soummer}, R., {Truong}, T., {Vasisht}, G., {Veicht}, A., {Vescelus}, F.,
  {Wallace}, J.~K., {Zhai}, C., \& {Zimmerman}, N. 2013, \apj, 768, 24

\bibitem[{{Pueyo} {et~al.}(2015){Pueyo}, {Soummer}, {Hoffmann}, {Oppenheimer},
  {Graham}, {Zimmerman}, {Zhai}, {Wallace}, {Vescelus}, {Veicht}, {Vasisht},
  {Truong}, {Sivaramakrishnan}, {Shao}, {Roberts}, {Roberts}, {Rice}, {Parry},
  {Nilsson}, {Lockhart}, {Ligon}, {King}, {Hinkley}, {Hillenbrand}, {Hale},
  {Dekany}, {Crepp}, {Cady}, {Burruss}, {Brenner}, {Beichman}, \&
  {Baranec}}]{Pueyo:2015}
{Pueyo}, L., {Soummer}, R., {Hoffmann}, J., {Oppenheimer}, R., {Graham}, J.~R.,
  {Zimmerman}, N., {Zhai}, C., {Wallace}, J.~K., {Vescelus}, F., {Veicht}, A.,
  {Vasisht}, G., {Truong}, T., {Sivaramakrishnan}, A., {Shao}, M., {Roberts},
  Jr., L.~C., {Roberts}, J.~E., {Rice}, E., {Parry}, I.~R., {Nilsson}, R.,
  {Lockhart}, T., {Ligon}, E.~R., {King}, D., {Hinkley}, S., {Hillenbrand}, L.,
  {Hale}, D., {Dekany}, R., {Crepp}, J.~R., {Cady}, E., {Burruss}, R.,
  {Brenner}, D., {Beichman}, C., \& {Baranec}, C. 2015, \apj, 803, 31

\bibitem[{{Rameau} {et~al.}(2013){Rameau}, {Chauvin}, {Lagrange}, {Boccaletti},
  {Quanz}, {Bonnefoy}, {Girard}, {Delorme}, {Desidera}, {Klahr}, {Mordasini},
  {Dumas}, \& {Bonavita}}]{Rameau:2013}
{Rameau}, J., {Chauvin}, G., {Lagrange}, A.-M., {Boccaletti}, A., {Quanz},
  S.~P., {Bonnefoy}, M., {Girard}, J.~H., {Delorme}, P., {Desidera}, S.,
  {Klahr}, H., {Mordasini}, C., {Dumas}, C., \& {Bonavita}, M. 2013, \apjl,
  772, L15

\bibitem[{{Sing} {et~al.}(2008){Sing}, {Vidal-Madjar}, {D{\'e}sert},
  {Lecavelier des Etangs}, \& {Ballester}}]{Sing:2008}
{Sing}, D.~K., {Vidal-Madjar}, A., {D{\'e}sert}, J.-M., {Lecavelier des
  Etangs}, A., \& {Ballester}, G. 2008, \apj, 686, 658

\bibitem[{{Skemer} {et~al.}(2012){Skemer}, {Hinz}, {Esposito}, {Burrows},
  {Leisenring}, {Skrutskie}, {Desidera}, {Mesa}, {Arcidiacono}, {Mannucci},
  {Rodigas}, {Close}, {McCarthy}, {Kulesa}, {Agapito}, {Apai}, {Argomedo},
  {Bailey}, {Boutsia}, {Briguglio}, {Brusa}, {Busoni}, {Claudi}, {Eisner},
  {Fini}, {Follette}, {Garnavich}, {Gratton}, {Guerra}, {Hill}, {Hoffmann},
  {Jones}, {Krejny}, {Males}, {Masciadri}, {Meyer}, {Miller}, {Morzinski},
  {Nelson}, {Pinna}, {Puglisi}, {Quanz}, {Quiros-Pacheco}, {Riccardi},
  {Stefanini}, {Vaitheeswaran}, {Wilson}, \& {Xompero}}]{Skemer:2012}
{Skemer}, A.~J., {Hinz}, P.~M., {Esposito}, S., {Burrows}, A., {Leisenring},
  J., {Skrutskie}, M., {Desidera}, S., {Mesa}, D., {Arcidiacono}, C.,
  {Mannucci}, F., {Rodigas}, T.~J., {Close}, L., {McCarthy}, D., {Kulesa}, C.,
  {Agapito}, G., {Apai}, D., {Argomedo}, J., {Bailey}, V., {Boutsia}, K.,
  {Briguglio}, R., {Brusa}, G., {Busoni}, L., {Claudi}, R., {Eisner}, J.,
  {Fini}, L., {Follette}, K.~B., {Garnavich}, P., {Gratton}, R., {Guerra},
  J.~C., {Hill}, J.~M., {Hoffmann}, W.~F., {Jones}, T., {Krejny}, M., {Males},
  J., {Masciadri}, E., {Meyer}, M.~R., {Miller}, D.~L., {Morzinski}, K.,
  {Nelson}, M., {Pinna}, E., {Puglisi}, A., {Quanz}, S.~P., {Quiros-Pacheco},
  F., {Riccardi}, A., {Stefanini}, P., {Vaitheeswaran}, V., {Wilson}, J.~C., \&
  {Xompero}, M. 2012, \apj, 753, 14

\bibitem[{{Skemer} {et~al.}(2014){Skemer}, {Marley}, {Hinz}, {Morzinski},
  {Skrutskie}, {Leisenring}, {Close}, {Saumon}, {Bailey}, {Briguglio},
  {Defrere}, {Esposito}, {Follette}, {Hill}, {Males}, {Puglisi}, {Rodigas}, \&
  {Xompero}}]{Skemer:2014}
{Skemer}, A.~J., {Marley}, M.~S., {Hinz}, P.~M., {Morzinski}, K.~M.,
  {Skrutskie}, M.~F., {Leisenring}, J.~M., {Close}, L.~M., {Saumon}, D.,
  {Bailey}, V.~P., {Briguglio}, R., {Defrere}, D., {Esposito}, S., {Follette},
  K.~B., {Hill}, J.~M., {Males}, J.~R., {Puglisi}, A., {Rodigas}, T.~J., \&
  {Xompero}, M. 2014, \apj, 792, 17

\bibitem[{{Soummer} {et~al.}(2011){Soummer}, {Brendan Hagan}, {Pueyo},
  {Thormann}, {Rajan}, \& {Marois}}]{Soummer:2011}
{Soummer}, R., {Brendan Hagan}, J., {Pueyo}, L., {Thormann}, A., {Rajan}, A.,
  \& {Marois}, C. 2011, \apj, 741, 55

\bibitem[{{Soummer} {et~al.}(2014){Soummer}, {Perrin}, {Pueyo}, {Choquet},
  {Chen}, {Golimowski}, {Brendan Hagan}, {Mittal}, {Moerchen}, {N'Diaye},
  {Rajan}, {Wolff}, {Debes}, {Hines}, \& {Schneider}}]{Soummer:2014}
{Soummer}, R., {Perrin}, M.~D., {Pueyo}, L., {Choquet}, {\'E}., {Chen}, C.,
  {Golimowski}, D.~A., {Brendan Hagan}, J., {Mittal}, T., {Moerchen}, M.,
  {N'Diaye}, M., {Rajan}, A., {Wolff}, S., {Debes}, J., {Hines}, D.~C., \&
  {Schneider}, G. 2014, \apjl, 786, L23

\bibitem[{{Soummer} {et~al.}(2012){Soummer}, {Pueyo}, \&
  {Larkin}}]{Soummer:2012}
{Soummer}, R., {Pueyo}, L., \& {Larkin}, J. 2012, \apjl, 755, L28

\bibitem[{{Yang} {et~al.}(2014){Yang}, {Apai}, {Marley}, {Saumon}, {Morley},
  {Buenzli}, {Artigau}, {Radigan}, {Metchev}, {Burgasser}, {Mohanty},
  {Lowrance}, {Showman}, {Karalidi}, {Flateau}, \& {Heinze}}]{Yang:2014}
{Yang}, H., {Apai}, D., {Marley}, M.~S., {Saumon}, D., {Morley}, C.~V.,
  {Buenzli}, E., {Artigau}, E., {Radigan}, J., {Metchev}, S., {Burgasser},
  A.~J., {Mohanty}, S., {Lowrance}, P.~L., {Showman}, A.~P., {Karalidi}, T.,
  {Flateau}, D., \& {Heinze}, A.~N. 2014, ArXiv e-prints

\bibitem[{{Zuckerman} {et~al.}(2011){Zuckerman}, {Rhee}, {Song}, \&
  {Bessell}}]{Zuckerman:2011}
{Zuckerman}, B., {Rhee}, J.~H., {Song}, I., \& {Bessell}, M.~S. 2011, \apj,
  732, 61

\end{thebibliography}
\end{document}